# Biomimetic Staggered Composites with Highly Enhanced Energy Dissipation: Design, Modeling, and Test

Pu Zhang[*], Mary A. Heyne, Albert C. To[†]

Department of Mechanical Engineering and Materials Science, University of Pittsburgh, Pittsburgh, PA 15261, USA

**Abstract**

We investigate the damping enhancement in a class of biomimetic staggered composites via a combination of design, modeling, and experiment. In total, three kinds of staggered composites are designed by mimicking the structure of bone and nacre. These composite designs are realized by 3D printing a rigid plastic and a viscous elastomer simultaneously. Greatly-enhanced energy dissipation in the designed composites is observed from both the experimental results and theoretical prediction. The designed polymer composites have loss modulus up to ~500 MPa, higher than most of the existing polymers. In addition, their specific loss modulus (up to 0.43 $Km^2/s^2$) is among the highest of damping materials. The damping enhancement is attributed to the large shear deformation of the viscous soft matrix and the large strengthening effect from the rigid inclusion phase.

---

[*] Email: p_zhang87@hotmail.com

[†] Corresponding author. Email: albertto@pitt.edu, Tel: 1-(412) 624-2052, Fax: 1-(412) 624-4846

# 1 Introduction

Damping is a fundamental and ubiquitous behavior of all solid materials (Lakes, 2009), while only a few solid materials reach the standard of engineering damping application. The design of high-performance wave or vibration absorbing structural components requires materials having high viscosity and moderate to high stiffness. The damping performance of materials (Lakes, 2009) is characterized by their complex modulus $E^* = E' + iE''$, with the real part $E'$ (storage modulus) and imaginary part $E''$ (loss modulus) being proportional to the storage energy and dissipated energy in the materials, respectively, and their ratio $\tan\delta = E''/E'$ known as the loss tangent (or viscosity) of materials. The loss modulus $E'' = E'\tan\delta$, a direct indicator of the energy dissipation, is also designated as the figure of merit of damping materials. In general, most materials show poor damping performance because they do not usually exhibit both high stiffness $E'$ and high viscosity $\tan\delta$ simultaneously (Lakes, 2002; Meaud et al., 2014). For example, soft polymers usually have high viscosity ($\tan\delta = 0.1\sim1$) but stiff materials like metals normally exhibit much lower viscosity ($\tan\delta < 0.001$) at room temperature (Lakes, 2009).

Figure 1 shows the loss modulus – mass density chart of several classes of materials, e.g. polymers, metals, alloys, ceramics, composites, biological materials, etc. Most of the material property data are drawn from the software package Granta (CES EduPack 2014) except InSn (Ludwigson et al., 2002), SiC-InSn (Ludwigson et al., 2002), ZnAl (Jaglinski and Lakes, 2012), SiC-ZnAl (Jaglinski and Lakes, 2012), bones (Freitas, 1999; Lakes, 2001; Smith Jr and Keiper, 1965), and the materials we tested including VeroWhitePlus (VW), digital polymer D9860, and staggered polymer composites. The ultimate goal of damping material design is to look for a material with either high loss modulus $E''$ to enhance energy dissipation or high specific loss modulus $E''/\rho$ to enhance damping and reduce weight at the same time. However, as shown in Fig. 1, few materials have loss modulus $E'' > 0.6\,\text{GPa}$ or specific loss modulus $E''/\rho > 0.5\,\text{K m}^2/\text{s}^2$. The materials with comparably high loss moduli include: 1) some soft metals or alloys such as lead, Mg, InSn, ZnAl, and their composites, 2) some non-technical porous ceramics like concrete and bricks (at high frequencies), and 3) biological materials like cortical bones (at quite low or high frequencies), plywood, and bamboo. Some hard materials, like hard metals, glasses, and technical ceramics, however, usually have quite low loss moduli. In addition, elastomers also have low loss moduli due to their low stiffness, although their viscosities are usually very high. Among all these materials in Fig. 1, polymers and foams are the most widely-used damping materials. In particular, polymers exhibit an intermediate loss modulus but comparably low density, which is favorable for lightweight component design. Foams are found to have quite low loss modulus but, surprisingly, their specific loss moduli $E''/\rho$ are on the same order as most other high damping materials. In brief summary, polymers, foams, and soft metals are the most popular conventional damping materials. However, newly developed damping composites, e.g. InSn-SiC, ZnAl-SiC, and polymer staggered composites, all have highly enhanced damping performance compared with conventional damping materials. It is expected that damping composites will replace the role of conventional damping materials in the future because composites can usually be designed and optimized to have performance better than single-phase materials.

Several methods have been proposed to design better performing damping materials, like introducing piezoelectric or magnetostrictive phases (Law et al., 1995; McKnight and Carman, 1999), employing phase transitions (Piedboeuf et al., 1998; San Juan et al., 2009), synthesizing nanocomposites (Suhr et al., 2005; Sun et al., 2009), and adding negative-stiffness phases (Jaglinski et al., 2007; Lakes et al., 2001). Nevertheless, one cannot underestimate the role of biological materials, especially those with high specific loss modulus, in inspiring and stimulating the design of materials with high energy dissipation. For example, it has long been known that foams mimic the microstructures of wood and bamboo (Gibson and Ashby, 1999; Meyers and Chen, 2014), while dissipative bio-inspired scaffolds have been recently synthesized by replicating the pore structure of cancellous bones (Porter et al., 2014). In addition, some theoretical and numerical works have shown that the bone- and nacre-like structure could be utilized to design phononic crystals with highly enhanced wave reflection/absorption performance (Chen and Wang, 2014; Yin et al., 2014; Zhang and To, 2013). Particularly, a recent theoretical work (Zhang and To, 2014) by the authors demonstrated that the bone-like staggered structure can be optimized to enhance the overall loss modulus significantly, which stems from the large shear deformation of the viscoelastic soft matrix. It would be of great significance to develop new composites with highly enhanced energy dissipation for engineering applications. Hence, the objective of this work includes two aspects: 1) Design and model both 2D and 3D staggered composites with better damping performance, and 2) Manufacture and test staggered polymer composites with enhanced energy dissipation.

## 2  Staggered Composite Design: from 2D to 3D

The 2D staggered composite shown in Fig. 2 mimics the microstructure of nacre and bone (Fratzl, 2008), which has drawn much attention and extensive investigation. For example, this model has been successfully used to explain the stiffness and toughness enhancement mechanism in 2D staggered composites (Barthelat and Rabiei, 2011; Jäger and Fratzl, 2000; Ji and Gao, 2004; Kotha et al., 2000; Kotha et al., 2001; Liu et al., 2006). In a 2D staggered composite, the hard prisms (or platelets) are dispersed in the soft matrix in a staggered manner (Fig. 2 (b)). The length and thickness of the prism is $l$ and $h$, respectively, which also define its aspect ratio as $\eta = l/h$ with $\eta \gg 1$. The thickness of the soft matrix layer is $h_c$ and the distance between two neighbor prism tips is $l_c$ ($l_c \ll l$). The loading-transfer characteristic in the 2D staggered composite is quite unique. The uniaxial loading along the longitudinal direction of prisms is mainly sustained by the shear deformation of the soft matrix within the shear region (Fig. 2 (b)) between two parallel prisms (Ji and Gao, 2004), even though the soft matrix in the tension region (Fig. 2 (b)) has a minor effect (Zhang and To, 2014). The force balance of the prism is illustrated in Fig. 2 (c), where $\tau_c$ is the shear stress in the shear region and $\sigma_c$ is the tensile stress in the tension region. Note that $\tau_c$ is not necessarily constant along the prism surface, especially when the aspect ratio $\eta$ is large. Since the loading-transfer is mainly induced by the shear region, we define the effective shear length of a prism as $l_s = l - l_c$ and the corresponding aspect ratio as $\eta_s = l_s / h$. Note that the difference between $\eta$ and $\eta_s$ is only significant when the aspect ratio $\eta$ is small.

3D staggered composites are designed as a substitute for the 2D model since the 2D model suffers from low loading-transfer ability. In reality, the staggered microstructure is akin to 3D rather than 2D in natural materials such as nacre and bone (Espinosa et al., 2009; Meyers et al., 2008; Reznikov et al., 2014). Compared with the 2D design, 3D staggered composites have more topological features to be designed and optimized, providing a greater possibility to leverage the loading-transfer ability. Therefore, two different kinds of 3D staggered composites are designed, which have square (Fig. 3 (a)) and hexagonal (Fig. 3 (d)) shaped prisms, respectively. The structure of 3D staggered composites is more complex than the 2D case. For both of the two types of 3D staggered composites, shown in Fig. 3, the prisms are distributed so that one prism's tip locates in the middle of adjacent prisms in the longitudinal direction. The prism's length, thickness, and aspect ratio are also designated as $l$, $h$, and $\eta$, respectively. Other parameters, like $l_s$, $\eta_s$, $h_c$, and $l_c$, can also be defined in accordance with the 2D case. The detailed arrangement of the prisms in 3D staggered composites is illustrated in Fig. 3 (b) and (e), which show the lattice structure of their transverse cross sections. The lattice points are indicated by black dots and defined by two lattice vectors $\boldsymbol{a}_1$ and $\boldsymbol{a}_2$. Similar to a crystal structure (Allen et al., 1999), the structure of a staggered composite is determined once its motif, the repeated unit cell resting at each lattice point, is prescribed. The representative motifs of 3D staggered composites are enclosed and highlighted by dotted closed circles in Fig. 3 (b) and (e). Each motif contains two arrays of prisms, which arrange in a staggered manner and are differentiated by solid lines and dashed lines in Fig. 3. The lattice points of the staggered composites can be obtained via translation operations based on the lattice vectors, which are defined as $\boldsymbol{a}_1 = [1\ 1], \boldsymbol{a}_2 = [\bar{1}\ 1]$, and $\boldsymbol{a}_1 = [\sqrt{3}\ 0], \boldsymbol{a}_2 = [0\ 1]$ for square and hexagonal cases, respectively. The edge length $a$ of the prism is derived from the geometric condition, where $a = h$ and $a = h/\sqrt{3}$ for square and hexagonal prism, respectively. Therefore, once the volume fraction of the hard phase is set as $\phi$, the soft layer thickness is determined by

$$h_c = \begin{cases} h(\phi^{-1} - 1) & \text{for 2D case} \\ h(\phi^{-0.5} - 1) & \text{for square or hexagonal prism} \end{cases} \quad (1)$$

It is found from Eq. (1) that the staggered composites with square and hexagonal prisms have equally thick soft matrix layers when $\phi$ and $h$ are fixed.

The loading-transfer characteristics are different for the two composites presented in Fig. 3. As shown in Fig. 3 (b), all of the four nearest neighbors of a square prism have different arrangement compared to itself, which induces shear stress $\tau_c$ on all of its four lateral surfaces (Fig. 3 (c)) once a uniaxial loading is applied to the composite. In contrast, the hexagonal prism (Fig. 3 (d) and (e)) has a different feature, that is, only four lateral surfaces out of six are subjected to shear stress loading, as illustrated in Fig. 3 (f). Thus, the shear region can be formed all around a square prism but only partially around a hexagonal prism. It is shown that both 3D designs have more effective loading-transfer ability than the 2D one. Additionally, similar to the 2D staggered composite, the tensile stress $\sigma_c$ is also be induced by the tension region, which has a minor effect on the deformation of a prism but should not be neglected when the aspect ratio $\eta$ of the prism is not large enough.

## 3 Theoretical Modeling

The complex modulus of staggered composites can be derived from the correspondence principle of linear viscoelasticity (Christensen, 1982; Lakes, 2009). Namely, the dynamic property of viscoelastic materials follows the same mathematical form as the elastic case by simply replacing all real elastic constants with complex values. Therefore, the elastic properties are derived first.

### 3.1 A Unified Shear-lag Model for Staggered Composites

A unified shear-lag model is presented to predict the overall elastic property of all three staggered composites. As shown in Fig. 4, the motif structure of each 3D composite is further reduced to a simple model (shaded area in Fig. 4) containing two reduced prisms bonded by a soft layer. The mechanical response of the reduced model is equivalent to the whole composite due to the lattice symmetry conditions. In fact, the reduced model in Fig. 4 is just a quarter of the Wigner-Seitz cell of the lattice structure for the two composites with square (Fig. 3 (b)) and hexagonal (Fig. 3 (e)) prisms. The cross sectional area $A$ of each reduced prism is

$$A = \begin{cases} \frac{1}{2}ah & \text{for 2D plane stress} \\ \frac{1}{4}a^2 & \text{for square prism} \\ \frac{3\sqrt{3}}{8}a^2 & \text{for hexagonal prism} \end{cases} \quad (2)$$

Note that the area $A$ is also able to be expressed as a function of $h$ only.

The shear lag model is illustrated in Fig. 5, which is composed of two reduced prisms and a soft layer between them. The maximum tensile stress $\sigma_m$ occurs in the middle of each prism due to the symmetry condition. Thereby, only a half of the reduced prism needs to be considered in the shear lag model. A local coordinate system is established at the center of the soft layer with $x$ denoting the longitudinal direction of the prism and hence $-l_s/4 \leq x \leq l_s/4$. In Fig. 5, prism 1 is subjected to tensile stress loading $\sigma_c$ and $\sigma_m$ on its two ends, where $\sigma_c$ is the tensile stress induced by the tension region. In contrast, prism 2 is subjected to the same tensile stress loading but on opposite ends. Suppose the displacement field in the prism is $u_j$ and $v_j$ along the $x$ and $y$ direction, respectively, with the subscript $j$ ($j = 1, 2$) indicating the prism number. An essential assumption of the shear lag model is that $u_j = u_j(x)$ and $v_j = 0$. Thus the prisms are in uniaxial tension and the misfit displacement $u_1 - u_2$ will induce shear deformation in the soft layer. Therefore, the shear strain $\gamma_c$ in the shear region of the soft layer is

$$\gamma_c = \frac{u_1 - u_2}{h_c} \quad (3)$$

In turn, the shear strain $\gamma_c$ asserts shear stress loading to the two prisms, as

$$\tau_c = \mu_c \gamma_c$$
$$= \frac{\mu_c}{h_c}(u_1 - u_2) \tag{4}$$

where $\mu_c$ is the shear modulus of the soft matrix.

The unknown displacements $u_1$ and $u_2$ will be solved from the equilibrium equation of the prisms, as (Gere and Timoshenko, 1984)

$$A \partial_x \sigma_1 = a\tau_c$$
$$A \partial_x \sigma_2 = -a\tau_c \tag{5}$$

where $A$ is shown in Eq. (2). Given that the tension strain is $\varepsilon_j = \partial_x u_j$ in the two prisms, the corresponding tensile stress is

$$\sigma_j = E_m \varepsilon_j$$
$$= E_m \partial_x u_j \tag{6}$$

where $E_m$ is the elastic modulus of the prism. Alternatively, the equilibrium equations in Eq. (5) can be further written in a displacement form by employing Eq. (6), as

$$\partial_{xx} u_1 - \frac{k^2}{2l_s^2}(u_1 - u_2) = 0$$
$$\partial_{xx} u_2 + \frac{k^2}{2l_s^2}(u_1 - u_2) = 0 \tag{7}$$

where $k$ is a dimensionless parameter related to the prism shape and material properties, as

$$k = \sqrt{\frac{2\mu_c a l_s^2}{E_m A h_c}} \tag{8}$$

It will be shown later that $k$ is a crucial geometrical parameter reflecting the loading-transfer ability of a staggered composite.

The boundary conditions of the two prisms are set as

$$\sigma_1 \big|_{x=-l_s/4} = \sigma_c$$
$$\sigma_1 \big|_{x=l_s/4} = \sigma_m$$
$$u_2 \big|_{x=-l_s/4} = 0$$
$$\sigma_2 \big|_{x=l_s/4} = \sigma_c \tag{9}$$

Up to now, the tensile stress $\sigma_c$ is still unknown, which is induced by the deformation of the tension region. It is seen from Fig. 6 that the rectangular tension region deforms into an octagon shape in the 2D staggered composite. Actually the deformation of the tension region is similar in 3D staggered composites. The tensile strain of the tension region between the two prism tips can be derived from the kinematic relation in Fig. 6, as

$$\varepsilon_c = \frac{2h_c}{l_c}\gamma_c\big|_{x=l_s/4} \tag{10}$$

Thus the average tensile stress $\sigma_c$ exerted on the prism tip is estimated to be

$$\sigma_c = \frac{c_c \varepsilon_c}{\phi} \tag{11}$$

where the term $\phi$ in the denominator is introduced to account for the tension effect of the soft material on the left and right sides of the tension region. The result of Eq. (11) has been proved to be valid even for 3D staggered composites. The term $c_c$ is the tensile stiffness (Sadd, 2014) of the soft material, as

$$c_c = \begin{cases} \dfrac{E_c}{1-v_c^2} & \text{for plane stress} \\ \dfrac{E_c(1-v_c)}{(1+v_c)(1-2v_c)} & \text{for other cases} \end{cases} \tag{12}$$

where $E_c$ and $v_c$ are the elastic modulus and Poisson's ratio of the soft matrix and $E_c = 2\mu_c(1+v_c)$. After substituting Eqs. (3) and (10) into Eq. (11), the tensile stress $\sigma_c$ is expressed in a displacement form, as

$$\sigma_c = \frac{2c_c}{\phi l_c}(u_1 - u_2)\big|_{x=l_s/4} \tag{13}$$

By substituting Eq. (13) into the boundary conditions in Eq. (9), the displacement field $u_j$ ($j=1,2$) is able to be solved from Eq. (7), as

$$u_j = \frac{\sigma_m}{2E_m}\left(\frac{l_s}{4} + x + \frac{1 + \frac{2c_c}{E_m l_c \phi}(\frac{l_s}{4}+x) + \frac{\zeta_j \cosh(kx/l_s)}{\cosh(k/4)}}{\frac{2c_c}{E_m l_c \phi} + \frac{k}{l_s}\tanh(k/4)}\right) \tag{14}$$

where

$$\zeta_j = \begin{cases} 1 & \text{for } j=1 \\ -1 & \text{for } j=2 \end{cases} \tag{15}$$

The tensile stress in the prism is determined by Eqs. (6) and (14), as

$$\sigma_j = \frac{\sigma_m}{2}\left(1 + \frac{\frac{2c_c}{E_m l_c \phi} + \frac{\zeta_j k \sinh(kx/l_s)}{l_s \cosh(k/4)}}{\frac{2c_c}{E_m l_c \phi} + \frac{k}{l_s}\tanh(k/4)}\right) \tag{16}$$

In addition, the shear stress distribution in the shear region of the soft matrix is obtained from Eqs. (4) and (14), as

$$\tau_c = \frac{\mu_c \sigma_m}{E_m h_c} \left( \frac{\frac{\cosh(kx/l_s)}{\cosh(k/4)}}{\frac{2c_c}{E_m l_c \phi} + \frac{k}{l_s}\tanh(k/4)} \right) \quad (17)$$

Equation (17) indicates that $\tau_c$ is not always constant along the prism. Only in the case that $k/4 \ll 1$ can $\tau_c$ be assumed to be constant.

### 3.2 Elastic Modulus of Staggered Composites

The overall elastic modulus of the staggered composites is derived based on the proposed shear lag model above. Bear in mind that the left end of prism 2 in Fig. 5 is fixed. Hence, the overall strain of the staggered composite is

$$\varepsilon = \frac{u_1|_{x=l_s/4}}{l_s/2} \quad (18)$$

On the other hand, the average stress in the composite is (Ji and Gao, 2004)

$$\sigma = (\sigma_m + \sigma_c)\phi \quad (19)$$

The overall elastic modulus $E = \sigma/\varepsilon$ of the staggered composite can be obtained from Eqs. (18) and (19) with given displacement field in Eq. (14) and stress distribution in Eq. (16). It is finally found that

$$\frac{1}{E} = \frac{1}{E_m \phi} + \frac{1}{c_c \frac{l_s}{l_c} + E_m \phi \frac{k}{4}\tanh(\frac{k}{4})} \quad (20)$$

where the first term on the right hand side shows the contribution from the prism and the second term represents the effect of the soft matrix. The simple form of Eq. (20) has a very strong physical implication, that is, the hard prism and soft matrix behave like a pair of springs in series while the soft matrix itself is just like connecting its tension region and shear region in parallel.

Note that Eq. (20) is a unified formula for both 2D and 3D staggered composites. The topology feature of different shapes of prisms is reflected in the parameter $k$ defined by Eq. (8). It is seen from Eq. (20) that the parameter $k$ is an effective stiffness indicator of the shear region of the soft matrix. A larger $k$ value indicates that the shear region is more effective in loading transfer. Equation (8) implies that, to increase $k$, one may either increase the edge length $a$ of the prism or reduce the soft layer thickness $h_c$, which can be achieved by changing the shapes of prisms. Specifically, $k$ has the following form for different staggered composites, as

$$k = \begin{cases} 2\eta_s \sqrt{\dfrac{\mu_c}{E_m(\phi^{-1}-1)}} & \text{for plane stress} \\ 2\sqrt{2}\eta_s \sqrt{\dfrac{\mu_c}{E_m(\phi^{-0.5}-1)}} & \text{for square prism} \\ \dfrac{4}{\sqrt{3}}\eta_s \sqrt{\dfrac{\mu_c}{E_m(\phi^{-0.5}-1)}} & \text{for hexagonal prism} \end{cases} \qquad (21)$$

It is found from Eq. (21) that the loading transfer ability of these staggered composites follows the sequence square > hexagonal > 2D. Therefore, the aspect ratio $\eta_s$ of the prism can be reduced if 3D designs are used instead of the 2D one.

The elastic modulus formula Eq. (20) derived from the shear lag model can be simplified when $k$ is small. This derives from the fact that $\tanh(k/4) \approx k/4$ when $k/4 \ll 1$. In this case, the simplified elastic modulus is

$$\frac{1}{E} \approx \frac{1}{E_m\phi} + \frac{1}{c_c l_s/l_c + E_m \phi k^2/16} \qquad (22)$$

After substituting the $k$ values in Eq. (21) to Eq. (22) and employing Eq. (12), the elastic modulus for the staggered composites can be written in a unified form, as

$$\frac{1}{E} \approx \frac{1}{E_m\phi} + \frac{f(\phi)}{\alpha \mu_c \phi \eta_s^2} \qquad (23)$$

where

$$f(\phi) = \begin{cases} 4(\phi^{-1}-1) & \text{for plane stress} \\ 2(\phi^{-0.5}-1) & \text{for square prism} \\ 3(\phi^{-0.5}-1) & \text{for hexagonal prism} \end{cases} \qquad (24)$$

and $\alpha$ is a correction factor introduced to account for the tension region effect, which is

$$\alpha = \begin{cases} 1+\dfrac{8h_c}{(1-\nu_c)\eta_s\phi l_c} & \text{for plane stress} \\ 1+\dfrac{4(1-\nu_c)h_c}{(1-2\nu_c)\eta_s\phi l_c} & \text{for square prism} \\ 1+\dfrac{6(1-\nu_c)h_c}{(1-2\nu_c)\eta_s\phi l_c} & \text{for hexagonal prism} \end{cases} \qquad (25)$$

Note that the effect of the tension region would be excluded once $\alpha = 1$. Equation (23) indicates that the effect of the tension region can be incorporated by simply multiplying the shear modulus $\mu_c$ of the soft matrix by a factor $\alpha$. In addition, this correction factor $\alpha \to 1$ when the shear aspect ratio $\eta_s$ increases, which indicates that the tension region effect is only significant when $\eta_s$ is small. It has been proven in our previous work (Zhang and To, 2014) that this correction factor can enhance the accuracy of the model a lot and should not be neglected.

The simplified elastic modulus formula in Eq. (23) is quite convenient to use and predicts satisfactory results in most cases, even though the one in Eq. (20) is more accurate in theory. For practical usage, one is strongly advised to check the criterion $k/4 \ll 1$ first with $k$ given by Eq. (21) to determine whether the shear-lag model or simplified model should be adopted.

### 3.3 Complex Modulus of Staggered Composites

Given that the elastic moduli of staggered composites have been derived in Section 3.2, their complex moduli can be readily obtained by replacing the elastic modulus $E_m$ of the hard phase and shear modulus $\mu_c$ of the soft matrix with the corresponding complex constants, $E_m^* = |E_m^*| e^{i\delta_m} = E_m' + iE_m''$ and $\mu_c^* = |\mu_c^*| e^{i\delta_c} = \mu_c' + i\mu_c''$, respectively. As a result, the parameter $k$ in Eqs. (8) and (21) and stiffness $c_c$ in Eq. (12) can also be transformed to their complex counterpart as $k^*$ and $c_c^*$ by replacing the material constants. Note that the Poisson's ratio $\nu_m$ and $\nu_c$ are assumed to be real and unchanged in both the static or dynamic cases. Finally, the complex modulus of staggered composites is obtained directly from Eq. (20) according to the correspondence principle, as

$$\frac{1}{E^*} = \frac{1}{E_m^* \phi} + \frac{1}{c_c^* \frac{l_s}{l_c} + E_m^* \phi \frac{k^*}{4} \tanh(\frac{k^*}{4})} \tag{26}$$

The exact storage modulus $E'$, loss modulus $E''$, and loss tangent $\tan\delta$ of staggered composites can all be obtained from Eq. (26) by separating the real and imaginary parts of $E^*$. In addition, the optimal damping situation can also be evaluated numerically by solving for the optimal aspect ratio $\hat{\eta}_s$.

Some approximated analytical formulae of $E^*$, $E'$, $E''$, and $\tan\delta$ for staggered composites are shown. These are derived based on the simplified model presented in Section 3.2. In this case, the complex modulus corresponding to Eq. (23) can be written as

$$\frac{1}{E^*} \approx \frac{1}{E_m^* \phi} + \frac{f(\phi)}{\alpha \mu_c^* \phi \eta_s^2} \tag{27}$$

Therefore, the storage modulus, loss modulus, and loss tangent are deduced from Eq. (27) directly by separating the real and imaginary parts of $E^*$, as

$$\begin{aligned}
E' &\approx \frac{|E_m^*| \alpha |\mu_c^*| \eta_s^2 \phi (|E_m^*| f \cos\delta_c + \alpha |\mu_c^*| \eta_s^2 \cos\delta_m)}{|E_m^*|^2 f^2 + \alpha^2 |\mu_c^*|^2 \eta_s^4 + 2|E_m^*| f \alpha |\mu_c^*| \eta_s^2 \cos(\delta_c - \delta_m)} \\
E'' &\approx \frac{|E_m^*| \alpha |\mu_c^*| \eta_s^2 \phi (|E_m^*| f \sin\delta_c + \alpha |\mu_c^*| \eta_s^2 \sin\delta_m)}{|E_m^*|^2 f^2 + \alpha^2 |\mu_c^*|^2 \eta_s^4 + 2|E_m^*| f \alpha |\mu_c^*| \eta_s^2 \cos(\delta_c - \delta_m)} \\
\tan\delta &\approx \frac{|E_m^*| f \sin\delta_c + \alpha |\mu_c^*| \eta_s^2 \sin\delta_m}{|E_m^*| f \cos\delta_c + \alpha |\mu_c^*| \eta_s^2 \cos\delta_m}
\end{aligned} \tag{28}$$

By employing the approximation in Eq. (27), the optimal aspect ratio $\hat{\eta}_s$ is obtained by evaluating $\partial E'' / \partial \eta_s = 0$, as

$$\hat{\eta}_s \approx \sqrt{\frac{|E_m^*| f \sin(\delta_c/2)}{\alpha |\mu_c^*| \sin(\delta_c/2 - \delta_m)}} \tag{29}$$

Note that $\alpha$ is a function of $\eta_s$ as given by Eq. (25). Hence the optimal aspect ratio $\hat{\eta}_s$ in Eq. (29) is governed by a quadratic function, which is easy to solve. In this optimal damping scenario, the optimal storage modulus $\hat{E}'$, optimal loss modulus $\hat{E}''$, and optimal loss tangent $\tan\hat{\delta}$ are expressed as

$$\begin{aligned}
\hat{E}' &\approx \frac{1}{2}|E_m^*|\phi \frac{\sin\delta_c}{\sin(\delta_c - \delta_m)} \\
\hat{E}'' &\approx |E_m^*|\phi \frac{\sin^2(\delta_c/2)}{\sin(\delta_c - \delta_m)} \\
\tan\hat{\delta} &\approx \tan(\delta_c/2)
\end{aligned} \tag{30}$$

It can be determined that the optimal loss modulus $\hat{E}''$ of staggered composites in Eq. (30) does not measurably change simply by tuning the prism shape. However, the optimal aspect ratio $\hat{\eta}_s$ in Eq. (29) can be adjusted by choosing different cross-sectional shapes for the prism. Interestingly, Eq. (30) indicates that the optimal loss modulus is achieved when the storage modulus $E'$ of the staggered composite is approximately half that of the Voigt bound (Christensen, 1982) and the phase delay $\delta$ is half that of the soft matrix. At this point, the staggered composite has both an intermediate stiffness and loss tangent, maximizing the loss modulus.

A special case occurs when the hard phase is purely elastic, i.e. $E_m^* = E_m$. This is quite useful for mineral based prisms, which are commonly seen in natural and synthesized staggered composites (Corni et al., 2012; Tang et al., 2003; Yao et al., 2011). In this particular case, the optimal aspect ratio, optimal storage modulus, and optimal loss modulus can all be further simplified to $\hat{\eta}_s \approx [E_m f/(\alpha |\mu_c^*|)]^{1/2}$, $\hat{E}' \approx 0.5 E_m \phi$, and $\hat{E}'' \approx 0.5 E_m \phi \tan(\delta_c/2)$, respectively. Note that similar formulae have been derived for a 2D staggered structure (Zhang and To, 2014) using another approach.

## 4 Manufacturing and Testing

The designed 2D and 3D staggered composites are manufactured by the 3D-printing technique to fabricate high-damping polymer composites and validate the proposed design and theory.

### 4.1 3D-Printing of Staggered Composites

The 3D-Printing technique enables rapid freeform fabrication of advanced structured materials and facilitates novel composites design (Babaee et al., 2013; Compton and Lewis, 2014; Dimas et al., 2013; Dimas and Buehler, 2014; Ge et al., 2013; Kang et al., 2014; Wang et al., 2011; Zhang et al., in press; Zheng et al., 2014). At present, the state-of-the-art PolyJet technique (developed by the Stratasys Ltd.) is able to manufacture designed parts with multiple photopolymers in a single job. Currently, a wide variety of photopolymers are provided, ranging from soft rubber-like materials to rigid plastics. This technique manufactures a part in such a way that the printer jets out photopolymer droplets based on a designed pattern first and then uses UV light to cure the polymer through photopolymerization. After one layer of droplets is cured, the printer proceeds to the next layer. The thickness of each layer is 16-30 microns. The designed staggered composites are manufactured by the Objet260 Connex 3D Printer with this PolyJet technique. The rigid photopolymer VeroWhitePlus (VW) is chosen for the hard prisms of the staggered composites. Meanwhile, the soft matrix is manufactured by the rubber-like digital material D9860, which is a mixture of the VW and TangoBlackPlus photopolymers. Figure 7 (a)-(c) show three examples of the manufactured staggered composites. The thickness of each prism is $h=1$ mm and the volume fraction of the prisms is $\phi=0.5$. The associated cross sections of the 3D staggered composites show regular arrangement of the prisms, which are consistent to the designed pattern shown in Figs. 2 and 3.

The manufactured staggered composites used for dynamic testing are shown in Fig. 7 (e)-(g). Rectangular cross sections are used for these specimens in order to be in compliance with the ASTM standards for dynamic testing of plastics (i.e. ASTM D4065 and ASTM D5026). Different from the staggered composite samples in Fig. 7 (a)-(c), two 25 mm long grip ends are printed using VW for each of the testing specimen, which will prevent damage to the D9860 material or staggered composites induced by the grips of the testing apparatus. The volume fraction of the VW prisms is $\phi=0.5$ for all the staggered composites printed. In addition, the thickness of each prism is $h=1$ mm while the aspect ratio $\eta$ varies from 6 to 18. The thickness of the soft matrix in the tension region is kept as $l_c=h$ so it is obvious that $\eta_s=\eta-1$ for these composite samples. All testing specimens are manufactured with their thickness direction normal to each printing layer and longitudinal axis along the printing head scanning direction. The specimens are kept on the building tray inside the printer chamber for two hours after printing in order to cool and further stabilize. The surfaces of the specimens are then cleaned by water jet to remove the support resin (SUP705). After removing the support resin, the specimens are dried in room condition for one hour and stored in specimen bags for ten hours before testing.

### 4.2 Static and Dynamic Testing of Constituent Materials

All tests are performed on an MTS880 system at room temperature $T=20\ ^\circ\text{C}$. The static tensile tests of VW and D9860 specimens are performed at a strain rate of $\dot{\varepsilon}=0.156\,\text{min}^{-1}$ and $\dot{\varepsilon}=0.178\,\text{min}^{-1}$, respectively. The dynamic mechanical behaviors are tested by applying a cyclic strain loading (at 1 Hz) on specimens and measuring the stress response. A pretension strain of $\varepsilon_0=1\%$ and $\varepsilon_0=11.6\%$, respectively, is applied to the VW and D9860 specimens before applying the cyclic strain loading. Note that the overall strain should not exceed the viscoelastic regime of the materials.

Figure 8 shows some representative testing results of the VW and D9860. For the static tensile responses, the VW has an elastic limit of $\varepsilon\approx 2\%$ while that for D9860 can be up to $\varepsilon\approx 45\%$. Some visible cracks would occur in the D9860 specimens once the tensile strain $\varepsilon>45\%$. The dynamic stress-strain relations of VW and D9860 are shown in Fig. 8 (b) and (d), respectively. The loss tangent of D9860 is nearly ten times larger than that of the VW. However, the energy dissipation capacity of D9860 is much less than the VW since it is too soft. Note that the loss modulus of D9860 is only 6% of the VW, which is proportional to the dissipated energy.

### 4.3 Dynamic Testing of Polymer Staggered Composites

The mechanical behaviors of the constituent materials used to fabricate the composites are introduced first. Table 1 shows the static and dynamic material properties of VW and D9860. It can be seen that the VW is a rigid plastic while the D9860 behaves like soft viscous rubber. The storage modulus of VW is a bit higher than its elastic modulus since the latter is measured at a quite low strain rate. The loss tangent $\tan\delta$ of VW is comparable to that of PMMA (at 1 Hz, room temperature). In contrast, the digital material D9860 is quite viscous with a loss tangent of $\tan\delta = 1.05$, a very high value for elastomers at room temperature. In addition, the storage modulus and loss modulus of D9860 are much higher than its elastic modulus.

The manufactured staggered polymer composites are tested under cyclic dynamic loading to validate the theory established above. The full set of the test specimens is shown in Fig. 7 (e)-(g). All specimens are tested at $T = 20\ °C$ with a frequency of 1 Hz. A pretension strain of ~1% is applied to the staggered composite specimens and then a cyclic strain loading at an amplitude of ~0.5% is used to perform the dynamic testing. Therefore, all tests are in the linear viscoelastic regime of the materials. The dynamic testing results of all three kinds of staggered composites are shown in Fig. 9. The theoretical prediction results are derived from Eq. (26) with material properties of constituent materials given in Table 1. It is seen that the theoretical values agree very well with the experimental results for the 2D staggered composites. Additionally, for the 3D staggered composites, the theoretical predictions are quite close to the experimental values, although some marginal difference is observed for the loss moduli prediction. There are two factors possibly affecting the results. First, the manufactured composite specimens are not identical to their designed CAD models, especially at the interfaces of the VW phase and D9860 phase, which will affect the overall material properties. Second, the theoretical model has made some simple assumptions for the deformation field in the shear regions and tension regions. Nonetheless, the accuracy of the theoretical prediction derived from the shear lag model is satisfactory for these staggered composites.

A comparison of the three staggered composites indicates that the two 3D staggered composites have higher loading-transfer ability than the 2D staggered composite. The optimal aspect ratio is $\hat{\eta} = 12$ (or $\hat{\eta}_s = 11$) for the two 3D staggered composites while that for the 2D composite is $\hat{\eta} = 29$. Thus, this proves that the 3D staggered composites are a more compact design for damping materials. The theoretical analyses show that the square prism has a slightly higher loading-transfer ability than the hexagonal prism. However, the experimental results indicate that the difference is negligible. In the optimal damping state for the 3D staggered composites, the storage modulus, loss modulus, and loss tangent are measured to be $E' = 520$ MPa, $E'' = 209$ MPa, and $\tan\delta = 0.4$, which are quite close to the optimal values predicted by Eq. (30), as $E' = 577$ MPa, $E'' = 247$ MPa, and $\tan\delta = 0.43$, respectively. In conclusion, these dynamic test results for staggered composite specimens validate the designed model and proposed theory above.

## 5  Discussions and Remarks

### 5.1  Simulation Results

To further understand the damping enhancement in staggered composites, some numerical examples are shown in Fig. 10, with emphasis on the effect of the hard phase type, aspect ratio $\eta_s$, and volume fraction $\phi$. Besides the theoretical results derived from Eq. (26), we also show the simulation results obtained from dynamic finite element analysis (FEA) to validate the theory by using the ANSYS software package (ANSYS® Academic Research, Release 14.5). The hard phase is considered to be purely elastic while the soft matrix is viscoelastic with $\tan\delta_c = 0.1$. Other material constants are set as $E_m = 1000\mu'_c$, $\nu_m = 0.27$, and $\nu_c = 0.4$. In addition, the thickness of the tension region is assumed to be $l_c = h_c$. The storage moduli, loss moduli, and loss tangents of three arrangements of staggered composites are shown in Fig. 10, which will be discussed below in order.

The storage modulus enhancement is similar to that of the elastic modulus, which has been studied extensively for 2D staggered composites in literature (Jäger and Fratzl, 2000; Ji and Gao, 2004; Kotha et al., 2000; Kotha et al., 2001). It was discovered that the soft material in the shear region exhibits large shear deformation, which stiffens the whole composite and enhances loading-transfer. The overall storage modulus $E'$ of the staggered composites increases rapidly when the aspect ratio $\eta_s$ increases, which eventually approaches the Voigt bound (i.e. $E' \to \phi E_m$) once $\eta_s \to \infty$ (Ji and Gao, 2004). A comparison of the storage modulus enhancement in these three staggered composites reveals their loading-transfer abilities, although a more rigorous analysis is to evaluate the geometrical parameters $k$ in Eq. (21). Given a certain value of aspect ratio $\eta_s$, the overall storage modulus $E'$ of the staggered composites is higher in the two 3D composites rather than the 2D case, implying higher loading-transfer abilities. This is easy to explain because the prisms in the 3D composites exhibit large contact areas with the soft matrix to transfer the shear stress. The 3D design requires a smaller aspect ratio $\eta_s$ than the 2D design, which provides feasibility to material synthesis and manufacturing since a shorter strengthening phase is usually favored.

The loss modulus enhancement in staggered composites differs from the storage modulus. The monotonic relation between $\eta_s$ and $E'$ does not exist for the loss modulus $E''$. Instead, $E''$ only increases to a certain maximum value of $\hat{E}''$ and then decreases gradually. This maximum point corresponds to a state with optimal damping performance and energy dissipation, which is the main focus of this work. Actually it is seen from Fig. 10 that $\hat{E}''$ of the whole composite is several orders of magnitude larger than the loss modulus $E''_c$ of the soft matrix, which is a significant amount of damping enhancement. Two questions are raised in regard to this optimal damping state from the application and manufacturing perspective, respectively. 1) How is the optimal loss modulus $\hat{E}''$ being achieved? 2) How can the optimal aspect ratio $\hat{\eta}_s$ be reduced to design more compact composites? Regarding the first question, it is found from the formula $\hat{E}'' \approx 0.5 E_m \phi \tan(\delta_c/2)$ that $\hat{E}''$ can only be increased by either increasing the elastic modulus $E_m$ of the hard phase, the volume fraction $\phi$, or the loss tangent $\tan\delta_c$ of the soft matrix. In other words, merely changing the shape of the strengthening phase will not increase $\hat{E}''$, which is confirmed by the results in Fig. 10. $\hat{E}''$ significantly varies with $\phi$ but is only slightly affected by the hard phase type. The second question is answered by the formula $\hat{\eta}_s \approx \sqrt{E_m f / (\alpha |\mu_c^*|)}$, which implies that $\hat{\eta}_s$ can be adjusted by either changing the volume fraction $\phi$ or the hard phase shape since the parameters $\alpha$ and $f(\phi)$ are related to them. The results in Fig. 10 show that $\hat{\eta}_s$ decreases whenever the volume fraction $\phi$ increases in the staggered composites. If the volume fraction $\phi$ is specified, the optimal aspect ratio $\hat{\eta}_s$ follows the sequence square < hexagonal < 2D in these staggered structures, which is exactly the inverse order of their loading-transfer abilities. Thus, it is concluded that the optimal aspect ratio $\hat{\eta}_s$ is smaller in a staggered structure with higher loading-transfer ability characterized by the geometrical parameter $k$ in Eqs. (8) and (21).

Distinct from $E'$ and $E''$, the loss tangent $\tan\delta$ decreases monotonically when $\eta_s$ increases, which implies that the staggered structures have smaller viscosity when $\eta_s$ is larger. This is because the staggered composite behaves more like its hard phase when $\eta_s$ increases. Therefore, the hard phase contributes more and more to the overall deformation, and $\tan\delta \to \tan\delta_m$ when $\eta_s \to \infty$. In contrast, the shear deformation of the soft matrix dominates the overall deformation when $\eta_s$ is quite small, in which case the staggered composite behaves like its soft matrix, and the loss tangent $\tan\delta \to \tan\delta_c$ when $\eta_s \to 0$. The optimal damping state corresponds to the case that the two phases have balanced deformation contribution to the whole composite, i.e. $\tan\delta = 0.5\tan\delta_c$, which is indicated by dotted lines in Fig. 10. Note that all the optimal states found are near these dotted lines since $\tan\hat{\delta} \approx \tan(\delta_c/2)$ according to Eq. (30).

## 5.2 Damping Enhancement Mechanism

Why does the loss modulus have an optimal value in staggered composites, and what is the mechanism? The answer is illustrated schematically in Fig. 11. The storage modulus $E'$ of the staggered composites is bounded by the Voigt bound, while its loss tangent $\tan\delta$ is bounded by that of the constituent materials. Consequently, the storage modulus $E'$ increases from $E'_c$ to $\phi E_m$ while the loss tangent $\tan\delta$ decreases from $\tan\delta_c$ to $\tan\delta_m$ when $\eta_s$ increases from 0 to ∞. Thus the loss modulus $E''$, i.e. the product of $E'$ and $\tan\delta$, might have an optimum $\hat{E}''$ at certain point $\eta_s = \hat{\eta}_s$ depending on the varying rate of the storage modulus and loss tangent. For the current problem, it is found that the loss modulus depends on the competition between the deformation in the hard phase and soft matrix. In the case that $\eta_s < \hat{\eta}_s$, the deformation of staggered composites is dominated by that of the soft matrix. The large shear deformation in the shear region of the soft matrix will increase the energy dissipation in the whole composite. However, once $\eta_s > \hat{\eta}_s$, large tension deformation occurs in the hard phase, which dominates the contribution from the soft matrix. Therefore, even though the overall storage modulus $E'$ of the composite still increases, $\tan\delta$ and $\hat{E}''$ decrease because the hard phase has low (or even none) energy loss. Therefore, the optimal state is achieved when the storage modulus contributions from the hard phase and soft matrix are equal. In this situation, the soft matrix provides the most efficient energy dissipation and both phases contribute to the overall storage modulus of the composite. It is the unique deformation mechanism transition from soft-matrix-dominant to hard-phase-dominant that induces the significant damping enhancement in staggered composites (Zhang and To, 2014). The storage modulus, loss modulus, and loss tangent at this optimal state can be found in Eq. (30). In addition, the optimal aspect ratio $\hat{\eta}_s$ is affected by the type of staggered composites.

In theory, a staggered composite with higher loading-transfer ability would exhibit smaller $\hat{\eta}_s$. Hence the 3D staggered composites with comparatively high volume fraction are recommended to design and manufacture better-performance damping materials.

### 5.3 Remarks on Damping Composite Design

Figure 12 shows the comparison of loss modulus enhancement in several composites comprising VW and D9860. The designed staggered polymer composites are found to have much higher energy dissipation than the Voigt and Reuss composites (Christensen, 1982). The optimal loss modulus of the staggered composite is only marginally higher than VW when the volume fraction is $\phi = 0.5$. However, the damping enhancement is much higher by increasing the volume fraction. For example, the loss modulus of the staggered polymer composites is up to ~500 MPa as the volume fraction of VW increases to > 90%, which is close to the upper limit of 600 MPa for most damping materials. In order to show the damping enhancement more clearly, the loss moduli and densities of staggered composites and the constituent materials, VW and D9860, are illustrated in the $E''$-$\rho$ chart in Fig. 1. It is found that the staggered polymer composites have much higher loss moduli $E''$ than most polymers. Even more remarkably, their specific loss moduli $E''/\rho$ (up to 0.43 $Km^2/s^2$) are among the highest ones of available damping materials, mostly because their comparably high damping but low density. The comparison indicates that the biomimetic staggered structure has great potential for the design and development of future high damping polymer composites.

Designing a composite with higher and higher energy dissipation ability is always desired for damping usage. Can the loss modulus of staggered composites be further increased? It has been mentioned that changing the shapes of the prisms does not change the optimal loss modulus much, except altering the optimal aspect ratio of prisms. However, it is seen from Eq. (30) that there are four ways to enhance the damping property further. (1) Increase the volume fraction $\phi$ of the hard phase. This is quite effective when the material properties cannot be changed. (2) Increase the modulus $|E_m^*|$ of the prisms, which enhances the storage and loss modulus of the whole composite. (3) Increase the loss tangent $\tan\delta_c$ of the soft matrix. This is actually quite crucial because the energy dissipation is largely a result of the viscosity of the matrix. (4) Increase the loss tangent $\tan\delta_m$ of the prisms. This has a minor effect compared to the former three methods. It is also suggested to use metal/ceramic materials as the hard phase and viscous polymers as the matrix so that the overall damping will be quite high.

## 6 Conclusions

In summary, the staggered composite design inspired from the bone and nacre structures has been proven to exhibit highly enhanced damping response. This originates from the unique deformation mechanism in the staggered composites. That is, the large shear deformation of the soft viscous matrix will result in high energy dissipation, while the hard phase endows the composite with high stiffness simultaneously. The optimal damping state is attained when the hard phase and soft phase have equal contribution to the overall storage modulus of the composites. According to this mechanism, three different kinds of staggered polymer composites are designed, tested, and compared. It is found that the 3D composites with square or hexagonal prisms have much higher loading-transfer ability than the 2D design. A much smaller aspect ratio of the hard phase is required to achieve the same storage modulus and loss modulus in the 3D design compared with the 2D one. The staggered composites have significantly increased loss modulus compared to their constituent materials and the Voigt and Reuss composites. It is found that the staggered polymer composites comprising VW and D9860 have loss moduli up to ~500 MPa, much higher than most polymers at room temperature. More impressively, their specific loss modulus, up to 0.43 $Km^2/s^2$, is among the highest ones of available damping materials. Hence, the biomimetic staggered structure is promising for the high-damping polymer composite design in the future.

## 7 Acknowledgements

The authors acknowledge the financial support from the Swanson School of Engineering at University of Pittsburgh.

# Tables

**Table 1.** Material properties of VeroWhitePlus (VW) and the digital material D9860. Dynamic material properties are measured at 1 Hz and 20 °C.

|       | $E$ (MPa)    | $\nu$ | $E'$ (MPa)  | $E''$ (MPa)  | $\tan \delta$  | $\rho$ (Kg/m$^3$) |
|-------|--------------|-------|-------------|--------------|----------------|-------------------|
| VW    | 1859 ± 11    | 0.33  | 2043 ± 80   | 215 ± 2      | 0.1~0.11       | 1160 ± 10         |
| D9860 | 2.1~2.2      | 0.45  | 12.1 ± 0.8  | 12.7 ± 0.3   | 1.05 ± 0.04    | 1145 ± 15         |

# Figure Captions

**Figure 1.** Loss modulus $E''$ versus mass density $\rho$ for various engineering and natural materials at room temperature. The same class of materials, e.g. polymers, metals and alloys, composites, etc., are indicated in the same color and enclosed together. The criterion of an excellent damping material is that either its loss modulus $E''$ or specific loss modulus $E''/\rho$ is high. Most of the materials have loss modulus $E'' < 0.6\,\text{GPa}$ and specific loss modulus $E''/\rho < 0.5\,\text{Km}^2/\text{s}^2$. The designed staggered polymer composites and their constituent materials (VW and D9860) are also shown. The oblique dashed lines are indicated for constant specific loss modulus $E''/\rho$ to assist lightweight damping material selection. Note that the damping properties of materials are frequency and temperature dependent.

**Figure 2.** Biomimetic design of 2D staggered composites from the bone structure. (a) The mineralized fibril structure of compact bones. The mineral prisms are arranged in a layer-wise staggered manner (Images (Fratzl, 2008) reprinted with permission of Nature Publishing Group). (b) Arrangement of the hard prisms in the designed composite. (c) Force balance diagram for an individual prism. The prism is subjected to shear stress loading on its lateral surfaces and uniaxial loading at its ends.

**Figure 3.** Schematic illustration of 3D staggered composites with (a)-(c) square prisms and (d)-(f) hexagonal prisms. (a) and (d) show the prism arrangement in each composite. (b) and (e) show the lattice structure of the transverse cross section of each composite. Lattice points are indicated by black dots. The motif is enclosed by a dotted closed circle, which contains two columns of prisms arranged in a staggered manner and being indicated by solid and dash lines, respectively. (c) and (f) illustrate the force balance diagram of a prism in the corresponding composite.

**Figure 4.** The reduced model (shaded area) for the motif structure of staggered composites to be used for the shear lag model. Each reduced model contains two reduced prisms and a soft layer between them, which is able to represent the structure and loading transfer characteristics of the whole composite. (a) Plane stress case. (b) Square prism. (c) Hexagonal prism.

**Figure 5.** The unified shear lag model for staggered composites. The model is applicable to both 2D and 3D staggered composites. The cross section illustrated here is for the reduced square prism but it can be generalized to other cases easily.

**Figure 6.** Schematic illustration of the deformation in the tension region of the soft matrix for a stretched 2D staggered composite. The initial boundary of the tension region is a rectangle, whereas the deformed boundary is an octagon. The tension deformation in the tension region can be generalized to 3D cases similarly.

**Figure 7.** Staggered polymer composites manufactured by the PolyJet 3D-printing technique by using two polymers of VW (in white color) and D9860 (in black color). The three kinds of staggered composites ($\phi = 0.5$, $\eta = 12$) are shown in (a) 2D composite, (b) 3D composite with square prisms, and (c) 3D composite with hexagonal prisms, all accompanied with their cross sections. (d)-(g) show the 3D-printed specimens for mechanical testing. All specimens contain two VW grip ends to prevent damaging induced by the grips of the load frame. (d) Testing specimens of VW (left) and D9860 (right). (e), (f), and (g) show a full set of 2D staggered composites, 3D staggered composites with square prisms, and 3D staggered composites with hexagonal prisms, respectively. Note that the seven specimens in (e)-(g) have a volume fraction of $\phi = 0.5$ and aspect ratios as $\eta$ = 6, 9, 10, 12, 14, 15, and 18, respectively (from left to right in each image).

**Figure 8.** Typical mechanical responses of the VeroWhitePlus (VW) photopolymer and the digital material D9860. (a) Static test of a VW specimen. (b) Cyclic dynamic test of a VW specimen at 1 Hz and 20 °C. (c) Static test of a D9860 specimen. (d) Cyclic dynamic test of a D9860 specimen at 1 Hz and 20 °C. The dissipated energy is equal to the area enclosed by the hysteresis circle (Lakes, 2009) in (b) and (d).

**Figure 9.** Storage moduli, loss moduli, and loss tangent of staggered composites obtained from theory (solid curves) and experiments (markers). (a) 2D staggered composite. (b) 3D staggered composite with squared prisms. (c) 3D staggered composite with hexagonal prisms. All dynamic tests are performed at 1 Hz and the room temperature.

**Figure 10.** Storage moduli, loss moduli, and loss tangent of different staggered composites with varying aspect ratio $\eta_s$ and hard phase content $\phi$. (a) 2D plane stress case. (b) 3D composite with square prisms. (c) 3D composite with hexagonal prisms. The optimal damping states occur near the dotted lines $\tan\delta = 0.5\tan\delta_c$.

**Figure 11.** Schematic illustration of the damping enhancement mechanism in staggered composites. The deformation mechanism changes from soft-matrix-dominate to hard-phase-dominate at the optimal damping state ($\eta_s = \hat{\eta}_s$). Both the storage modulus and loss tangent have monotonic relations with respect to the aspect ratio $\eta_s$ while the loss modulus has a peak.

**Figure 12.** Comparison of the loss modulus enhancement in several composites comprising polymers VW and D9860 with different volume fraction $\phi$. The Voigt and Reuss composites are compared with the current staggered composites.

**Figure 1**

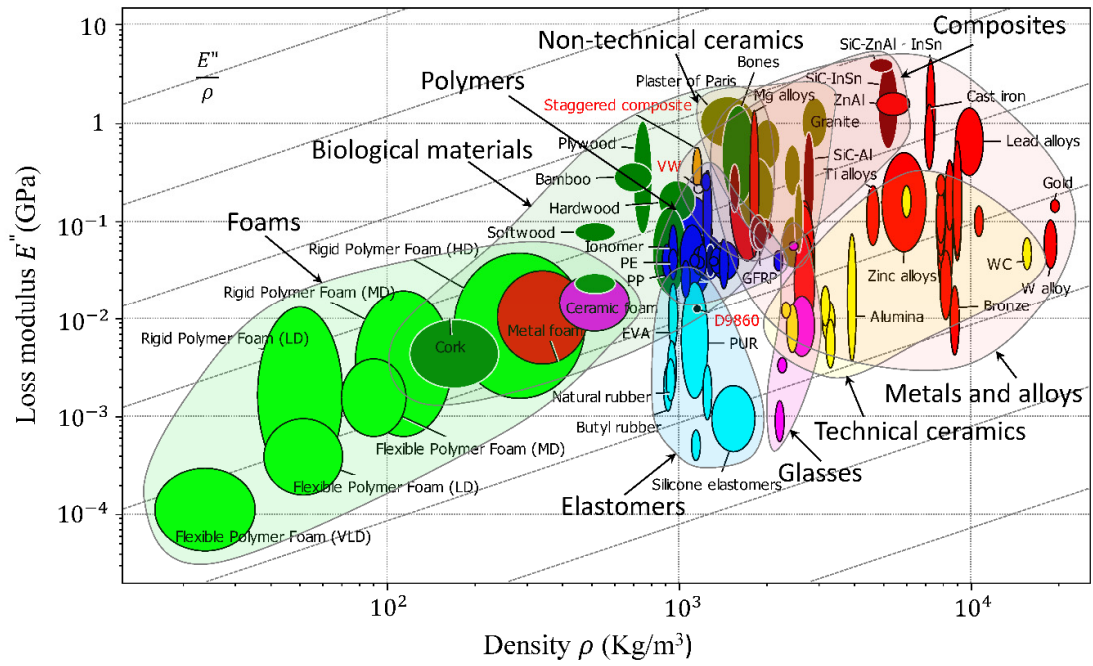

**Figure 2**

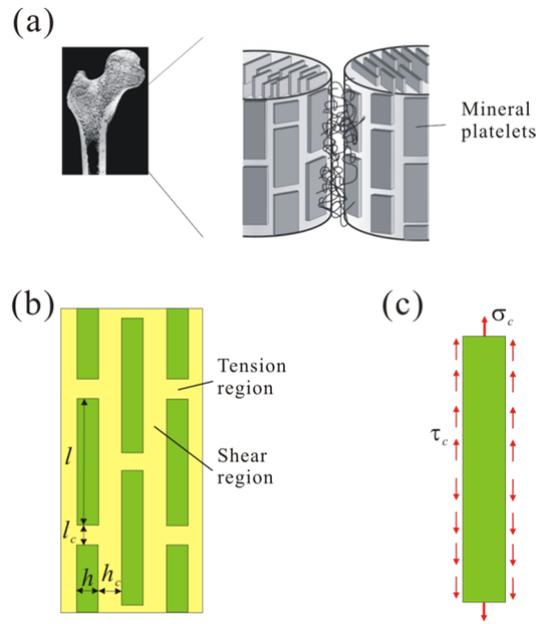

**Figure 3**

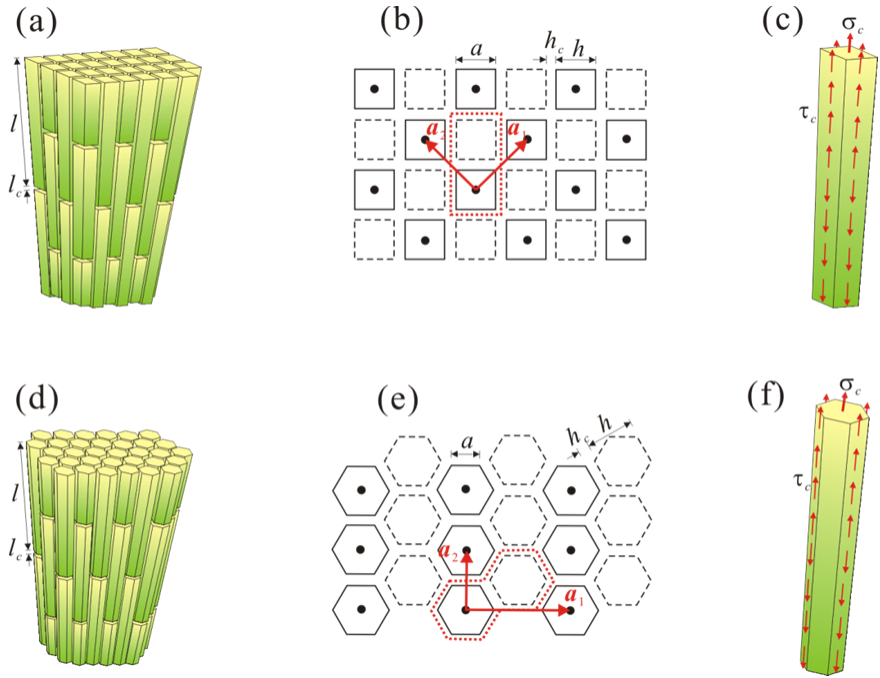

**Figure 4**

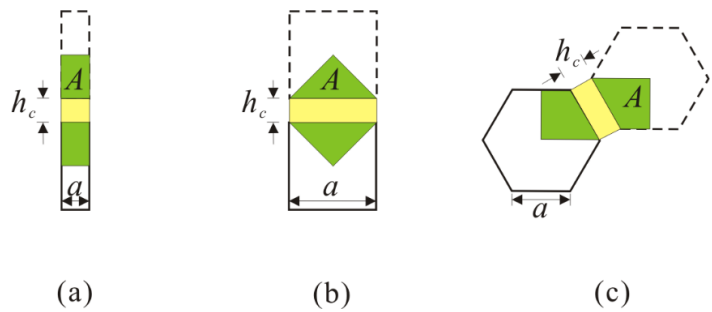

(a) (b) (c)

**Figure 5**

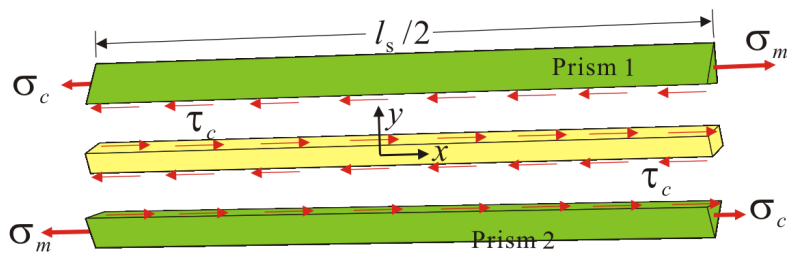

**Figure 6**

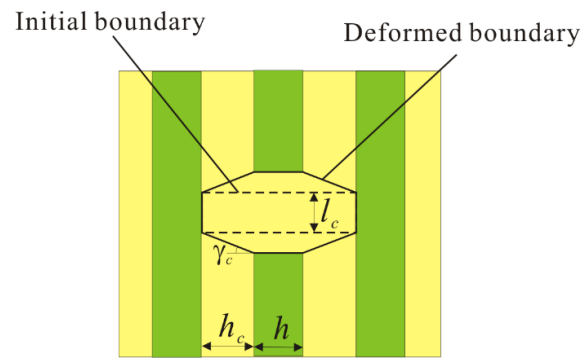

**Figure 7**

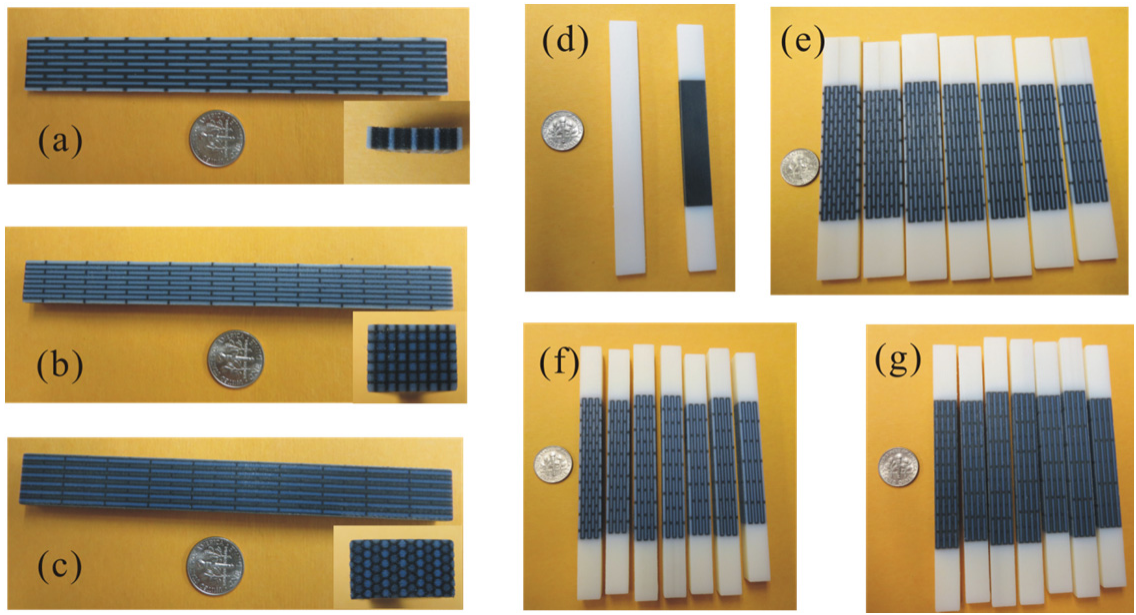

**Figure 8**

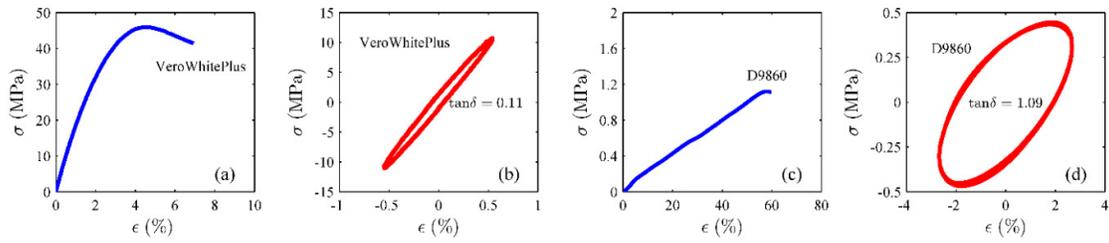

**Figure 9**

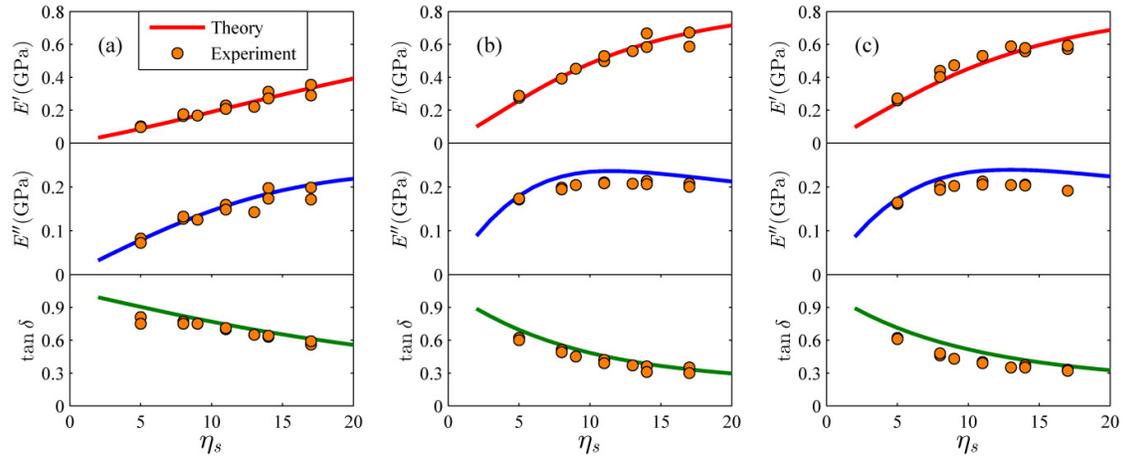

**Figure 10**

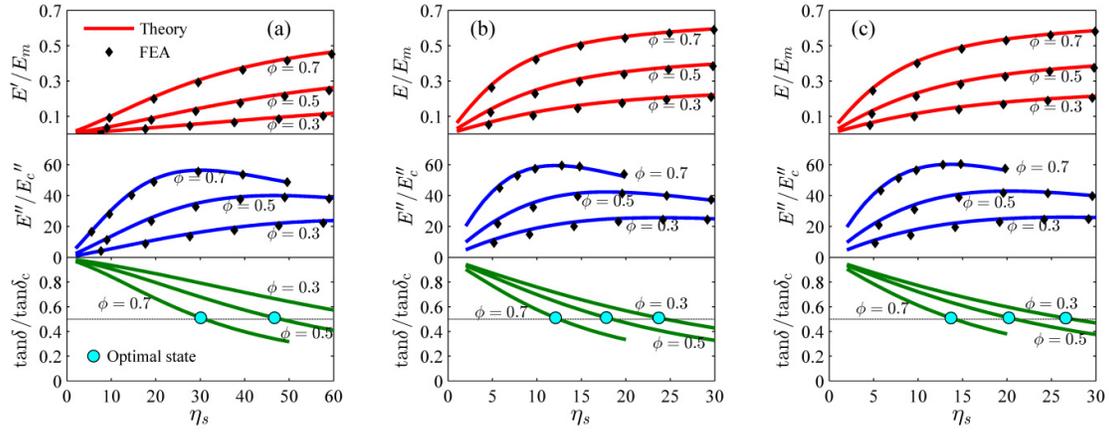

**Figure 11**

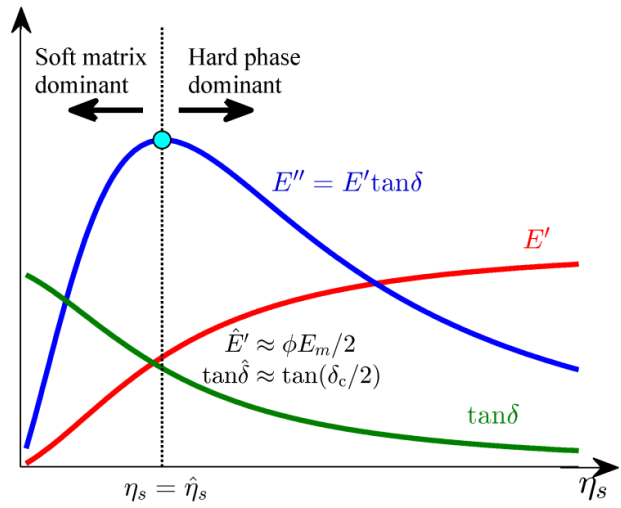

**Figure 12**

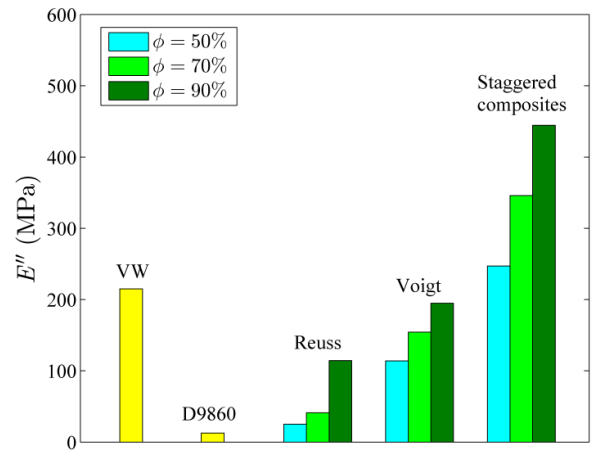